\documentstyle[twocolumn,prl,aps]{revtex}

\input epsf

\begin{document}

\bibliographystyle{unsrt}

\twocolumn[\hsize\textwidth\columnwidth\hsize\csname
@twocolumnfalse\endcsname

\title{Mesoscale magnetism at the grain boundaries in colossal
magnetoresistive films}

\author{Yeong-Ah Soh$^1$, G. Aeppli$^1$, N. D. Mathur$^2$, and
M. G. Blamire$^2$}

\bigskip

\address{$^1$NEC Research Institute, 4 Independence Way, Princeton, NJ
08540, $^2$Department of Materials Science, University of Cambridge,
Cambridge CB2 3QZ, UK}

\date{\today}
\maketitle
\begin{abstract}
We report the discovery of mesoscale regions with distinctive magnetic
properties in epitaxial La$_{1-x}$Sr$_{x}$MnO$_{3}$ films which
exhibit tunneling-like magnetoresistance across grain boundaries.  By
using temperature-dependent magnetic force microscopy we observe that
the mesoscale regions are formed near the grain boundaries and have a
different Curie temperature (up to 20 K {\it higher}) than the grain
interiors.  Our images provide direct evidence for previous
speculations that the grain boundaries in thin films are not
magnetically and electronically sharp interfaces.  The size of the
mesoscale regions varies with temperature and nature of the underlying
defect.
\end{abstract}
] 

Since the observation of large low-field magnetoresistance in
polycrystalline La$_{1-x}A_{x}$MnO$_{3}$ ($A={\rm Ba, Ca,
Sr}$)\cite{ju143sh,schiffer336sh,hwang041p,gupta629}, where the effect
was attributed to domain wall scattering or spin-polarized tunneling
(SPT) between grains, much attention has been drawn to the role of
grain boundaries (GBs) in the magnetotransport of
manganites\cite{evetts593p,balcells697sh,klein371sh}. To isolate the
GB contribution to magnetotransport, mesoscopic devices were patterned
on La$_{1-x}A_{x}$MnO$_{3}$ films grown on bicrystal substrates with
an artificial GB\cite{mathur266p,steenbeck968sh,isaac038p}.  Despite
evidence that the GBs contribute in a crucial way to the electrical
properties of colossal magnetoresistance (CMR)
materials\cite{jin413sh}, there is no microscopic information on the
magnetic and electronic properties of the GBs themselves. Here, we
provide the first such information in the form of temperature ({\it
T})-dependent images obtained by magnetic force microscopy (MFM),
which has a much higher resolution (30 nm) than other magnetic
microscopies used to study
manganites\cite{lecoeur934p,petrov061p,lin014p}.  The images lead to
the discovery of mesoscale regions around the GBs which have magnetic,
and therefore also electronic, properties different from those away
from the grain GBs.  Apart from yielding an essential fact about thin
transition metal oxide films, which are important both scientifically
and technologically (especially where microelectronic applications are
concerned), our work is significant as a demonstration of the use of
force microscopy for discovering a spatially inhomogeneous {\it
T}-dependent magnetic phenomenon.

We prepared our samples via the same procedures used to prepare the
material for the GB magnetotransport
devices\cite{mathur266p,isaac038p}. Epitaxial
La$_{1-x}$Sr$_{x}$MnO$_{3}$ films with $x = 0.3$ and 0.23 were grown
by pulsed laser deposition on bicrystal SrTiO$_{3}$(001) substrates
with an artificial GB where the crystals are misaligned by
$45^{\circ}$.  On one side of the artificial GB the [100] crystal axis
of SrTiO$_{3}$ crystal is parallel to the GB, whereas on the other
side, the [100] axis is rotated by $45^{\circ}$ with respect to the
GB. The films are 100 nm thick.  Because the lattice constant of
SrTiO$_{3}$ is larger than that of La$_{1-x}$Sr$_{x}$MnO$_{3}$, the
films grown on SrTiO$_{3}$ are subject to tensile strain, resulting in
a suppression of $T_{c}$ compared to the bulk\cite{tsui421sh}
and a magnetization vector {\it M} lying in the plane of the
film\cite{kwon229p}. We confirmed the in-plane orientation of {\it M}
by measuring hysteresis loops at 300 K with the field {\it H} in the
plane of the sample using a SQUID magnetometer (see the inset of
Fig.~\ref{MvsT}).  {\it M} as a function of {\it T} was measured for
the film composition $x=0.3$ and its corresponding target material
(Fig.~\ref{MvsT}).  $T_{c}$ for the film and the target material are
350 K and 373 K, respectively, with the target material displaying a
$T_{c}$ which is 23 K higher than that of the film.

We used MFM to image the magnetic domain patterns in our films. 
The microscope was operated in the tapping mode where the phase shift
$\phi$ of the oscillating cantilever was detected as the tip was
scanned at a fixed height above the sample.  The magnetic tips were
magnetized along the long axis of the tips, which is perpendicular to
the film plane.  All the scans were done for zero external field.
Since the films have an in-plane easy axis, the MFM will be sensitive
to the regions where the magnetization vector is rotating, or - in
other words - the magnetic domain walls (DW).  Panel {\bf B} of
Fig.~\ref{afmmfm} shows a typical magnetic domain pattern for the
La$_{0.77}$Sr$_{0.23}$MnO$_{3}$ film around the artificial GB.  Panel
{\bf A} is the corresponding atomic force microscope (AFM) image,
which attests both to the sharpness of the DW as well as the
smoothness ($\lesssim 1$ nm root mean square (RMS) variation in
thickness over a $20 ~\mu{\rm m} \times 20 ~\mu{\rm m}$ area) of the
film. We can clearly see a sharp magnetic DW that coincides with the
artificial GB.  In addition, there are magnetic DWs on opposite sides
of the GB which are not nucleated along any feature visible in the AFM
image. Although they meet at the artificial GB, they have different
orientations on either side, which represent the crystal orientation
of the underlying substrate and consequently that of the film itself.
On the left side of the GB, the DWs are parallel or perpendicular to
the GB.  From the right side of the GB, a magnetic DW initially
emanates at an angle of $45^{\circ}$ relative to the GB.  Because a
$45^{\circ}$ angle also characterizes the rotation of the crystal axes
across the GB in the bicrystal substrate, this confirms that our
La$_{0.77}$Sr$_{0.23}$MnO$_{3}$ film is epitaxial and a bicrystal with
an artificial GB coincident with that of the bicrystal substrate.  The
magnetization vectors of the magnetic domains in the film are coupled
to the crystal axes of the substrate - as we cross the GB the
magnetization vector has to rotate, and therefore it is natural to
form a magnetic DW at the GB.  Fig.~\ref{afmmfm} shows large magnetic
domains at room temperature (RT).  Their sidelengths are of order
$50~\mu{\rm m}$, much larger than the 60 nm film thickness.

We established the evolution of magnetic domains as a function of {\it
T} for an $x=0.3$ bicrystal film by imaging the domains using the MFM
with the sample mounted on a variable {\it T} sample
stage\cite{soh607}.  
As shown in the left column of Fig.~\ref{MFMvsT}, raising {\it T}
towards $T\simeq T_{c}$ (panel {\bf B}) reduces the magnetic contrast
which exists at RT (panel {\bf A}).  As we increase {\it T} further,
above $T_{c}$, we notice a remarkable thing - the emergence of a new
magnetic region (indicated by blue in panel {\bf C}), very different
from the domain pattern observed below $T_{c}$.  Specifically, at
355K, there is a distinct mesoscale region along the GB, with a
half-width of approximately $0.7~\mu{\rm m}$.  At $T=360$~K, the
mesoscale region shrinks to a half-width of $0.5~\mu{\rm m}$ and it
disappears entirely at $T=370$~K.  The effect is observed not only at
the artificial GB (which was introduced intentionally) but also at
other locations on the film (see right column of Fig.~\ref{MFMvsT}),
where there are naturally occurring substrate defects.  These defects
can be clearly seen in the topography (AFM) channel.  In such a
location, we followed the evolution of the magnetic images in smaller
{\it T} increments. The {\it T} dependence of the new magnetic region
is the same as that around the artificial GB.  It appears at $T\gtrsim
T_{c}$, peaks at 360 K, and vanishes at 370 K.

To quantify the magnetic contrast, we calculated the RMS of $\phi$ in
the MFM image at each {\it T}, where $\phi_{rms}={\sqrt \langle
(\phi(x,y)-\phi_{av})^2 \rangle}$.  $\phi(x,y)$ represents the phase
shift at each pixel and $\phi_{av}$ is the average phase shift in the
MFM image.  The results are plotted in Fig.~\ref{varqvsT}{\bf C}.  The
magnetic contrast between the mesoscale region (indicated by blue in
panels {\bf C}, {\bf D}, {\bf G}, and {\bf H}) and the rest of the
sample (indicated by yellow and green in {\bf C}, {\bf D}, {\bf G},
and {\bf H}) arises because the mesoscale region is ferromagnetic (FM)
and the rest is paramagnetic (PM) at $T\gtrsim T_{c}$.  When both the
mesoscopic region and the bulk part of the film are FM (when $T< T_c$,
i.e. at 300 K), the magnetization vector has a large magnitude, and
therefore the interaction between the tip and the sample is strong,
resulting in a large magnetic contrast in the MFM image at regions
where the magnetization vector rotates (i.e., at the DWs).  When the
film becomes weakly FM (when $T \sim T_{c}$), the interaction between
the tip and the sample is weak all across the sample.  Therefore, the
variation of $\phi$ across the scanned image is small resulting in a
small $\phi_{rms}$.  As the ferromagnetism of the bulk part
diminishes, the difference in magnetic force between the mesoscopic
region (FM) and the rest of the sample (PM) increases as shown in
Fig.~\ref{varqvsT}{\bf C}.  At $T > T_{c}$, the regions away from the
GBs become PM, whereas the regions near the GBs stay FM.  There is
then a large difference in the magnetic force experienced by the tip
in the two regions
, which gives rise to a large variation of $\phi$ across the image,
resulting in a large $\phi_{rms}$.  Eventually, the difference in
magnetic force between the two regions decreases and vanishes as the
ferromagnetism of the mesoscopic region vanishes and the whole film
becomes PM.  

It is notable that the new magnetic regions above $T_{c}$ are
magnetized in one direction when imaged with the MFM (as evidenced by
the dominant blue color in the images) and the force between the tip
and these FM regions is always attractive.  We believe that the tip
magnetizes these regions in a direction perpendicular to the film
plane as it scans the sample.  This is likely to happen given that
these regions are soft magnets with a small coercive field ($H_{c}
\lesssim 8$~G), as described below and shown in Fig.~\ref{varqvsT}{\bf
B}.  From the images, $T_c$ can be mapped spatially by locating the
FM-PM boundary as a function of $T$ since at the boundary $T=T_c$
(Fig.~\ref{varqvsT}{\bf A}).  That the mesoscopic regions shrink as
{\it T} is raised shows that $T_c$ varies spatially, with the regions
closer to the grain boundary having a higher $T_c$.

We have discovered that a thin manganite film has inhomogeneous
magnetic properties due to both natural and artificial GBs. Because
the GBs, especially when the micron scale healing length observed
directly in the MFM images is taken into account, occupy an
appreciable volume of the film, we expect to see evidence for
ferromagnetism extending up to 370 K in very sensitive bulk
magnetization measurements. Therefore, we have measured hysteresis
loops using SQUID magnetometry.  Indeed, we found that although the
{\it T} dependence of the order parameter indicates a $T_{c}= 350$ K
for $x=0.3$ film composition, there is still a tiny hysteresis left at
$T\gtrsim T_{c}$ (see Fig.~\ref{varqvsT}{\bf B}).  The tiny hysteresis
loops close exactly at the same temperature ($T=370$~K) at which the
mesoscale regions disappear in the MFM image.  The hysteresis loops at
$T\gtrsim T_{c}$ have the same shape whether the field is applied in
the plane or perpendicular to the plane of the film.  This indicates
that the FM regions around the GBs are isotropic, in contrast to the
rest of the film, which shows an in-plane easy axis throughout the
whole {\it T} range including the vicinity of $T_c$.

We also measured magnetization loops in the bulk powder (ceramic)
target ${\rm La_{0.7}Sr_{0.3}MnO_{3}}$ from which the film was grown.
Hysteresis ceases at 370 K, which is also the Curie temperature
deduced from {\it M} versus {\it T} data (see Fig.~\ref{MvsT}).
Therefore, the FM regions that exist at the GBs at $T> T_c$ in the
film do not exist in the bulk ceramic sample.  If we compare the
magnetization per unit volume for the film to that of the bulk powder
target for temperatures above $T_c$ of the film, we can estimate the
volume fraction of the film which is FM at these temperatures.  At
$T=355$, 360, and 365 K, which are intermediate between the nominal
Curie temperatures of the film and the powder, the FM volume fractions
are 0.03, 0.02, and 0.01, respectively. These values are consistent
with an estimate ($0.02=1~ \mu$m width/$50~ \mu$m distance between
GBs) of the volume fraction occupied by mesoscale FM regions in the
film, and so indicate that the small ``foot'' in $M(T)$ above $T_{c}$
for the film (see Fig.~\ref{MvsT}) is actually due to GB magnetism.
Almost needless to say, in the absence of our MFM data, the foot would
have many possible interpretations.

We attribute the variation of local $T_{c}$ to the local variation of
strain in the film. The effect of strain on the $T_{c}$ of CMR films
has been reported by various groups\cite{millis588,gan978sh,rao794p}
and is by now a well-accepted phenomenon.  Depending on the lattice
mismatch between the film and the substrate, the strain on the film
can be modulated, which in turn modulates $T_{c}$ substantially.  This
phenomenon has been attributed to the Jahn-Teller distortion arising
from biaxial strain\cite{millis588}. The coincidence of Curie
temperatures and the magnetization per unit volume make it very likely
that the mesoscale regions in our films are very similar to the bulk
starting material. Therefore, near the GBs, the film is strain
relieved leading to a $T_{c}$ almost the same as that of the bulk.  On
the other hand, away from these crystal imperfections, the film is
under tensile strain and its $T_{c}$ is suppressed by 20 K.  The width
of the mesoscale region is an indication of the range over which
strain propagates from the GB.

To summarize, we have discovered distinctive magnetic properties -
most notably a higher Curie temperature - in mesoscale regions around
GBs in manganites. The distinctive properties obtain whether the GBs
occur naturally in unplanned fashion, or are introduced deliberately
via a bicrystal substrate. They therefore need to be incorporated in
descriptions of the electronic transport in all CMR films.  Instead of
sharp boundaries that divide crystal grains, one needs to consider
separate regions around the GBs, which may have not only different
magnetic properties, but also different electronic structures. A
magnetically disordered region or mesoscale region around the GBs has
been invoked\cite{gupta629,evetts593p,klein371sh} to explain
magnetotransport results, which could not be explained by SPT at a
sharp interface.  Our results represent the first direct evidence that
the interfaces have magnetic properties which are actually modulated
over mesoscopic distances.  Beyond its implications for a topic of
great current interest, namely exploiting GBs for electronic devices,
our experiment is significant because it is pioneering in the sense of
imaging a spatially varying Curie temperature and using force
microscopy as a quantitative tool in the study of a {\it T}-dependent
magnetic phenomenon.

\acknowledgements{We are very grateful to Peter Littlewood for helpful
discussions and the contacts which made this collaboration possible,
as well as to Chang-Yong Kim for valuable X-ray analysis of our
films.}


\begin{figure} 
\caption{{\it M} versus {\it T} of La$_{0.7}$Sr$_{0.3}$MnO$_{3}$ film
and bulk powder sample.  For the film, {\it H} was 20~G ($>$ coercive
field) and was applied in the plane of the sample and perpendicular to
the grain boundary.  For the bulk powder sample which had a needle
shape, {\it H} was 500 G and was applied along the long axis of the
sample.  The $T_{c}$ of the film is 350 K, which is 23 K lower than
the $T_{c}=373$ of the powder target material.  The inset shows the
magnetic hysteresis loop of the film at 300 K, with the field applied
in the plane of the film and perpendicular to the grain boundary.}
\label{MvsT} 
\end{figure}

\begin{figure}
\caption{Room temperature AFM and MFM images of the region around the
artificial grain boundary in the La$_{0.77}$Sr$_{0.23}$MnO$_{3}$ film
grown on a bicrystal SrTiO$_{3}$ (001) substrate.  The scan size is 74
$\mu$m for both images. {\bf A} The AFM image shows the topography of
the film, which indicates the presence of an artificial grain
boundary. The z scale is in units of nm.  {\bf B} The MFM image
displays the magnetic domain walls in the system, one of which
coincides with the artificial grain boundary.  The z scale represents
the phase shift of the oscillating cantilever in degree units.}
\label{afmmfm}
\end{figure}

\begin{figure}
\caption{Evolution of magnetic pattern in the
La$_{0.7}$Sr$_{0.3}$MnO$_{3}$ film around the artificial grain
boundary and around natural defects as a function of {\it T}.  Images
{\bf A}, {\bf B}, {\bf C}, and {\bf D} were taken around the
artificial grain boundary at 300, 350, 355, and 365 K,
respectively. The scan size for these images was 5 $\mu$m and a common
z scale was used, which is displayed at the left of the images in
degree units.  Images {\bf E}, {\bf F}, {\bf G}, and {\bf H} display
the magnetic pattern around natural defects at 300, 350, 355, and 365
K, respectively.  The scan size for this region was 14 $\mu$m.  The z
scale for the scans is displayed at the right of the image in degree
units.  As {\it T} is raised from 300 K, the magnetic contrast
diminishes.  As {\it T} is raised above $T_{c}$, a new magnetic region
emerges around the grain boundary.  The width and strength of this new
magnetic region evolves as a function of {\it T}. The region
eventually vanishes at $T=370$ K.}
\label{MFMvsT}
\end{figure}

\begin{figure} 
\caption{Comparison between magnetic information derived
from MFM measurements and standard bulk magnetometry data. {\bf A}
Dependence of $T_c$ on distance from artificial grain boundary, as
established from the MFM images in Fig.~\ref{MFMvsT}.  {\bf B}
Magnetic hysteresis loops of the La$_{0.7}$Sr$_{0.3}$MnO$_{3}$ film
for $T>T_{c}$ were measured with the magnetic field applied in the
plane of the film and perpendicular to the artificial grain boundary.
The loops, which are open at $T < 370$~K, close at 370 K.  {\bf C}
Left axis shows the magnetic contrast of the MFM images in
Fig.~\ref{MFMvsT} represented by the RMS of $\phi$ in the MFM signal
as a function of {\it T}.  This is compared against the quantity
plotted on the right axis, which is the coercive field of the
La$_{0.7}$Sr$_{0.3}$MnO$_{3}$ film at $T > T_{c}$ extracted from the
magnetic hysteresis loops in panel {\bf B}.  The two quantities show a
similar {\it T} dependence.  The solid line is a smooth fit of the RMS
of $\phi$ and serves as a guide to the eye.}
\label{varqvsT}
\end{figure}

\end{document}